\def\jAJ{{\em AJ\/}}
\def\jAL{{\em Astr.Lett.\/}}
\def\jAPJ{{\em ApJ\/}}
\def\jAPJL{\jAPJ{ \em (Lett.)\/}}

\def\jMN{{\em MNRAS\/}}
\def\jPASP{{\it PASP\/}}
\def\jNature{{\it Nature\/}}

\newcommand{\RomanNum}[1]{\@Roman{#1}}
\documentstyle{mn}

\title[The Orbital Period of HDE226868/Cyg X-1]
{The Orbital Period of HDE226868/Cyg X-1}
\author[LaSala et al.]{
J. LaSala$^{1,2}$, P.A. Charles$^{1}$, R.A.D. Smith$^{1}$,
M. Ba\l uci\'nska-Church $^{3}$, and M.J. Church $^{3}$\\
$^{1}$ University of Oxford, Department of Astrophysics, 
   Keble Road, Oxford OX1 3RH\\
$^{2}$ University of Southern Maine, Physics Department, Portland, 
Maine 04104-9300\\
$^{3}$ School of Physics \& Astronomy, University of Birmingham, 
   Birmingham B15 2TT\\
}

\begin{document}

\maketitle


\begin{abstract}

\noindent
We present epoch 1996, high-quality radial velocity data for HDE 226868, the
optical counterpart of Cygnus X-1.  Combining our results with all published
historical data, we have derived a new ephemeris for the system of
HJD2450235.29 $+ n{\times}5.5998$, which allows accurate orbital phase
calculations to be made for any X-ray observations over the last 30 years.
We find no evidence for any period change as has been suggested by Ninkov,
Walker \& Yang (1987). We discuss the shortcomings of previous work in
establishing the period and orbital elements.

\end{abstract}


\section{Introduction}

Cygnus X-1, identified with the bright (V$\sim$8) star HDE226868 (Bolton
1972; Webster \& Murdin 1972), has long been regarded as the best black hole
candidate among the high-mass X-ray binaries.  As such, it has been an
object of extensive observation over the past two and a half decades.
Perhaps surprisingly, there remain some important uncertainties and
discrepancies in the derived properties of the system.  In particular,
Ninkov, Walker, \& Yang (1987, hereafter NWY) report evidence of possible
period variation, discussed below, and additional periodicities on
timescales ranging from 39 days to 4.5 years have been suggested by Kemp,
Herman, \& Barbour (1978), Wilson and Fox (1981), Priedhorsky, Terrell, \&
Holt (1983), and Walker and Quintanilla (1978).\\

HDE226868 is a single-lined spectroscopic binary.  The optical spectrum has
been classified O9.7Iab by Walborn (1973), with variable emission at HeII
$\lambda$4686 and, less prominently, at the Balmer lines. The radial
velocity period of 5.6 days has been established for over two decades,
and the definitive orbital elements and ephemeris were published by Gies and
Bolton (1982, hereafter GB).  They give a period of 5.59974 $\pm$ 0.00008
days, a precision which translates to an uncertainty of only $\pm$0.02 cycle
in phase at the present time.\\

However, NWY, in a subsequent radial-velocity study, reported a somewhat longer
period of 5.6017 $\pm$ 0.0001 days, and suggest the possibility of a period
increase with time.  These results have the effect that the orbital phase of
the Cyg X-1 system is then highly uncertain at the present epoch. The phase
calculated using the ephemeris of NWY differs from the GB phase by 0.5
cycle, and if the period is in fact varying, the phase at the present epoch
is completely indeterminate.\\

If correct, a time-varying period would have important implications for the
mass transfer rate and hence evolution of the binary.  Furthermore, with
this indeterminacy in phase, it is impossible to ascribe and interpret
accurately features in the X-ray light curve such as the dipping behaviour,
in particular the distribution of dipping with orbital phase (e.g. Remillard
and Canizares 1984). To resolve these ambiguities we therefore undertook a
program of observations to re-establish the orbital ephemeris of
HDE226868. These data also allowed us to undertake a more detailed
investigation of the long-term stability of the orbit, as high quality data
now exist over a $\sim$30 year baseline.  Our new ephemeris has thereby
allowed us to complete a survey of the distribution of X-ray dipping with
orbital phase (Balucinska-Church et al. 1998). \\

\section{Observations}

Through the La Palma Service Programme, and with the assistance of regularly
scheduled observers, we obtained CCD spectra of HDE226868 using the
Intermediate Dispersion Spectrograph on the Isaac Newton Telescope on 17
nights in May and June 1996.  These included 11 consecutive nights covering
two complete orbital cycles.  In all we obtained 37 spectra of the object,
each covering a spectral range of 4100 to 4900\AA, with a dispersion of
approximately 0.8 \AA/pixel.  Exposures varied from 100 to 200 seconds, and
yielded a S/N in excess of 100.  In addition, we observed 19 Cep, a radial
velocity standard of similar spectral type, on two nights, for reference and
calibration purposes.  \\

The spectra were extracted and reduced using standard IRAF routines.  The
extracted spectra were wavelength calibrated using a Cu-Ar arc recorded
immediately before and after the program spectra, and then normalized. Our
wavelength calibration accuracy had rms residuals of $\sim$0.10\AA. Typical
HDE226868 and 19 Cep spectra are shown in Figure 1. \\

\section{Period Determination}

We determined radial velocities by cross-correlation of the HDE226868
spectra with the 19 Cep spectra, using IRAF routines.  Separate
determinations were made for the hydrogen and helium lines.  We used the HeI
absorption lines at 4388, 4472, 4713 and 4921\AA. The resulting velocities
are given in Table 1, and were reduced to heliocentric velocities using the
catalogue value (Hoffleit 1982) for the radial velocity of 19 Cep.  We
performed period searches on both sets of velocities, as well as on our
velocities in combination with previously determined velocities from the
literature.  We used several different period-search techniques, including
Scargle periodograms, Fourier searches, and chi-square minimization to a
circular orbit fit.  All methods yielded similar results, as follows.  \\

The best fit period to the hydrogen line velocities for our data alone is
5.566 $\pm$ 0.012 days, while the best fit to our helium line data is 5.629
$\pm$ 0.015 days.  The apparent discrepancy here is startling, as is the
amount of apparent change in period from the canonical value in either case.
Both of these issues can be readily resolved.  \\

It is well established that the hydrogen lines are contaminated by emission
whose velocity is in approximate antiphase with the absorption lines (GB;
Hutchings, Crampton, \& Bolton 1979).  This produces variable and irregular
line profiles which yield spurious velocities and periods when
cross-correlated with standard spectra.  So any period determined from the
hydrogen line velocities is highly suspect.  For this reason we reject all
velocity determinations based on the hydrogen lines, and in working with
historical data below, we use only published velocities determined for the
helium lines.  \\

The fact that the best-fit period to the helium lines seems to be $2\sigma$
larger than previously determined values should also be viewed with some
scepticism.  In fact, with a dense grid of observations over a relatively
short period of time, the quality of fit is good over a much broader period
range than suggested by the quoted uncertainty.  Fitting our data to the NWY
period, or to the GB period, produces $\chi^2$ residuals which are not
significantly worse than those for our best fit.  In fact, the formal
uncertainty calculated by the fitting routines is undoubtedly too small;
this is because the scatter of the data is such that even for the best
orbital models the resulting $\chi^2$ is too large for the formal
uncertainty calculations to be valid.  We shall see below that this is not
unique to our results but in fact has been a problem for most, and perhaps
all, previous published orbital solutions for Cyg X-1; uncertainties have
been systematically underestimated, resulting in apparent discrepancies
where none really exist.  \\

To test for the possibility of period variation, we combined our
observations with all previously published helium-line velocity data
(Brucato and Zappala 1974; GB; NWY) and performed a $\chi^2$ minimization
weighted fit to a circular orbit both with a fixed period and with a period
which varies linearly with time, using the MINUIT package of
function-fitting routines (James 1994).  The results are summarized in Table
2 and plotted in figure 2.  The best fixed period is 5.5998 $\pm$ 0.0001
days, with a $\chi^2$ of 4284; the best variable period is 5.5997 $\pm$
0.0001 days, with $\dot{P}$ of 3.8 $\times$ 10$^{-7}$ and $\chi^2$ of 4231.
But with only 216 data points, this means that the $\chi^2_\nu$ is $\sim$20
in both cases.  Our new period agrees well with the photometric period
derived recently by Voloshina, Lyutyi \& Tarasov (1997) and a solution from
HeI $\lambda$6678 measurements by Sowers et al (1998).  Our values for
$\gamma$ and K are within 2$\sigma$ of previous results (NWY and GB).\\

The variable-period model produces a slightly better fit to the data, but,
again, with a $\chi^2$ difference of less than 1.5\%, the improvement in
quality of fit between that and the fixed-period model is {\it not}
statistically significant.  We conclude that there is no compelling evidence
for period variation in Cygnus X-1.  We have therefore adopted the orbital
elements of the best-fit constant period model in all further discussion.  \\

We computed O--C residuals for all the helium-line velocities and performed
period searches on these.  No significant periodicity was found in these
residuals.  In particular, we find no evidence of the suggested
periodicities at 39d or 78d (Kemp, Herman, \& Barbour 1978), or 4.5 years
(Wilson and Fox 1981) or of the 294 day X-ray modulation (Priedhorsky,
Terrell, \& Holt 1983) or the 91d photometric period (Walker and Quintanilla
1978).  The absence of a 294 d signal was also noted 
by Gies \& Bolton (1984).\\

\section{Discussion}

Our best-fit period for the combined datasets, 5.5998 $\pm$ 0.0001 d is
identical to well within the standard errors with the GB period of 5.59974
$\pm$ 0.00008 d, and our T$_0$, or epoch of inferior conjunction, agrees
with the prediction of the GB ephemeris to within 0.014 phase, or 0.078 d.
Since the uncertainty in the periods yields an uncertainty in the GB
prediction of T$_0$ of $\pm$ 0.12 days at this epoch, this is excellent
agreement.  Thus our results confirm the ephemeris of GB and refute
suggestions of period variation by NWY.  \\

Why, then, did NWY's result show such a discrepancy and suggest a change in
period?  They reported that the period determined from their data alone was
5.60172 $\pm$ 0.00003 d, differing by 20 $\sigma$ from the period determined
from all previous observations.  We suggest several sources for this error.

First of all, we performed our own period searches on the data published by
NWY; the best fit we find is at 5.6002 d, or only 2 $\sigma$
greater than the GB value, so we may be looking at much ado about a
typographical error.  In addition, there is excellent phase agreement
between the NWY data and the GB ephemeris, again suggesting no real period
discrepancy.                                                          \\

Finally, it seems that the error estimates for both the NWY period and their
period determined from historical data are too small by as much as an order
of magnitude.  This is primarily because the formal error calculations used
to determine these error estimates are based on the assumption of a very
good fit of the model to the data, i.e. that $\chi^2_\nu\sim$1. This
condition is {\it not} met by any of the data sets and fits used in this
work, presumably because the estimates of the uncertainties in the
velocities were too small, although perhaps because of variability in the
source.  We recalculated the fits, trying larger estimates for the velocity
uncertainties until the reduced chi-square criterion was satisfied.  We
found that the period uncertainties then were about an order of magnitude
larger than reported by the original authors.  For example, treating NWY's
data in this way yields a best-fit period of 5.6002 $\pm$ 0.0003 d, and
the discrepancy with GB disappears.  \\

An additional source of underestimated error uncertainty is the inclusion by
both NWY and Bolton (1975) of two velocities obtained by Seyfert and Popper
(1941), ostensibly to improve the precision of the period determinations.
For example, Bolton (1975) uses these points to decrease his uncertainty
estimate by a factor of 10; NWY do not discuss the effect of including these
points on their uncertainty.  Popper (1996, personal communication) suggests
that these velocities are ``very weak reeds on which to hang significant
conclusions'', the velocities having considerable uncertainties and being
based on averaging velocities from lines of several different species.
Because of the problems with hydrogen-line velocities discussed above,
inclusion of these points is therefore unlikely to improve the period
uncertainty.\\

GB also introduced ``velocity corrections'' to many of their velocities,
shifting spectra to fit a mean interstellar K-line velocity due to
instabilities in their spectrograph.  It is not unlikely that this
introduced uncertainties larger than the mean errors they report.\\

\section{Conclusions}

We have carried out a programme of radial velocity determinations for
the black hole binary Cyg X-1, and have combined these data with data
used previously to determine the period, to provide a new orbital
ephemeris for the source. A critical consideration of the errors 
associated with previous work has shown these to be underestimated.
Based on this, our main conclusion is that there is no evidence for a
change in orbital period as suggested by Ninkov et al. Finally, our
new ephemeris allows the orbital phase calculation for Cyg X-1
with an error that is much reduced compared with the error that would
be attached to extrapolating the Gies and Bolton ephemeris with its
quoted accuracy to the present epoch.\\

\section*{Acknowledgements}

We are very grateful to the La Palma support astronomers, and particularly
Don Pollacco, who operate the Service Programme, and to those astronomers
who allowed the Cyg X-1 observations to be taken during their own time. The
Isaac Newton Group of telescopes is operated on the island of La Palma by
the Royal Greenwich Observatory in the Spanish Observatorio del Roque de Los
Muchachos of the Instituto de Astrof\'\i{}sica de Canarias. JLS wishes
gratefully to acknowledge the following sources of partial support for his
participation in this work: the Margaret Cullinan Wray Charitable Lead
Annuity Trust (through the AAS Small Grants Program); the NASA JOVE program;
a Theodore Dunham, Jr., Grant of the Fund for Astrophysical Research; the
International Astronomical Union Exchange of Astronomers Program; and the
Department of Astrophysics of the University of Oxford.

\begin{table}\caption{Radial Velocity Data for HDE226868.
\label{tab:dyn}}

\begin{center}

\begin{tabular}{ccccccc}

 Heliocentric&Phase&  \multicolumn{2}{c} {Hydrogen Lines} & \multicolumn{2}{c} {Helium 
Lines}&  \\
   Julian Date &&    $V_r$ &  $\sigma_v$ &    $V_r$ &  $\sigma_v$ & {0 - C} \\
   2450000+ && \multicolumn{2}{c} {km s$^{-1}$} &\multicolumn{2}{c} {km
s$^{-1}$} & {km s$^{-1}$}  \\ \\

     228.739&   0.830& -71.6&   6.8&   -54.4&   5.2&    17.1\\  
     228.742&   0.831& -74.8&   7.2&  -59.0&   5.3&     12.3\\  
     228.744&   0.831& -75.5&   7.5&   -56.2&   5.8&    15.1\\  
     229.715&   0.004& -5.6&    6.6&    12.4&    7.0&   17.8\\  
     229.724&   0.006& -7.6  &  6.6   &  10.3    &6.1&  12.8 \\ 
     230.731&   0.185& 73.7  &  7.1   &  86.2    &5.6&  23.1\\ 
     230.733&   0.186& 69.5  &  6.9   &  88.0    &5.5&  24.0\\ 
     231.736&   0.365& 55.6  &  5.0   &   64.9   & 4.4& 13.7 \\ 
     231.738&   0.366& 52.8    & 4.8   & 61.8 &    4.4& 10.9 \\  
     232.736&   0.544& -33.8 &   6.9  & -34.2  & 5.8&   -8.2 \\  
     232.738&   0.544& -35.9 &   6.4  &-36.9  & 5.8 &   -10.9\\  
     233.723&   0.720& -75.2  & 7.0  & -64.6 &   3.9 &  14.9\\  
     233.726&   0.721& -80.7  & 7.2  &-68.8 &   4.5 &   10.8\\  
     234.730&   0.900& -53.2 &   6.0  &-43.7 &   5.0 &  6.2\\  
     234.733&   0.901& -52.4 &   6.0  &-39.9 &   5.3 &  9.4\\  
     235.726&   0.078& 25.1 &    7.2  & 47.4 &    4.1 & 17.3\\  
     235.730&   0.079& 26.6 &    7.1  &  46.5 &    5.0& 15.9 \\  
     236.737&   0.258& 69.3 &    7.1  & 80.6  &   6.9 & 10.7\\  
     236.740&   0.259& 68.7 &    6.8  & 77.0  &   7.3 & 6.7\\  
     237.738&   0.437& 12.0    & 7.9  & 7.4  &    5.6 & -16.3\\  
     238.745&   0.617& -64.5   & 5.5  & -60.6 &   3.4 & -4.6\\ 
     238.746&   0.617& -69.0   & 7.0  & -65.2 &  4.2 &  0.0\\ 
     238.748&   0.618& -71.4 &  6.9 &  -75.4 &   5.2 &  -19.1\\ 
     241.722&   0.149& 51.4  &  5.5 &  60.8  &  4.0 &   5.4\\ 
     242.704&   0.324& 54.8  &  5.3 &  60.0   & 4.0 &   -2.1\\  
     242.705&   0.324& 56.7  &  5.9 &  61.6    & 3.4 &  -0.5\\  
     243.721&   0.506& -2.8  &  5.4 &  2.9     & 4.9 &  11.1\\  
     243.722&   0.506& -4.9  &  5.0 &  -7.8    & 2.8 &  0.4\\  
     253.726&   0.292& 72.0 &   9.8 &  49.5 &    5.9 &  -17.9\\  
     253.727&   0.292& 74.2   & 10.2 &44.1   & 5.4 &    -23.4\\  
     254.721&   0.470& 8.4 &     9.9 &  -7.3 &    6.3 & -16.0\\  
     254.721&   0.470& 12.0    & 10.3 & -7.9   & 6.3 &  -16.6\\  
     255.673&   0.640& -64.6  & 6.3 & -70.6   & 5.0 &   -7.1\\  
     255.674&   0.640& -63.2 &   6.3 &  -67.0 &   6.5&  -3.5 \\ 
     255.675&   0.640& -68.0   & 5.9 & -71.1  & 4.9 &   -7.5\\

\end{tabular}     
\end{center}
\end{table}

\begin{table}\caption{Orbital Solutions for HDE226868.
\label{tab:dyn}}

\begin{center}

\begin{tabular}{lcc} 

& Variable Period Model       &    Fixed Period Model \\ \\

$P_0$ (days) & 5.5997(1) &5.5998(1) \\
\.{P} & 3.8(4) X 10$^{-7}$  & 0 (fixed) \\
$T_0$ (JD 2,440,000+) & 1869.10(5) & 10235.29(1) \\
$\gamma$ (km s$^{-1}$) & -5.4(1) & -5.4(1) \\
K (km s$^{-1}$) & 75.53(15) & 75.48(15)\\
$\chi^2$  & 4231 & 4283\\

\end{tabular}
\end{center}
{\footnotesize
Figure in parentheses is uncertainty in last digit.}
\end{table}

\bigskip

{\bf Figure 1:} Typical rectified spectra of HDE226868 (top) and the radial
velocity standard 19 Cep (bottom) obtained with the La Palma 2.5m INT.  The
spectral types are very similar, but note the HeII $\lambda$4686 emission in
HDE226868.

\bigskip

{\bf Figure 2:} Best fitting fixed period, circular orbit to all helium
radial velocity data (ours plus all previously published material, see
text).  The lower panel shows the residuals to the fit.

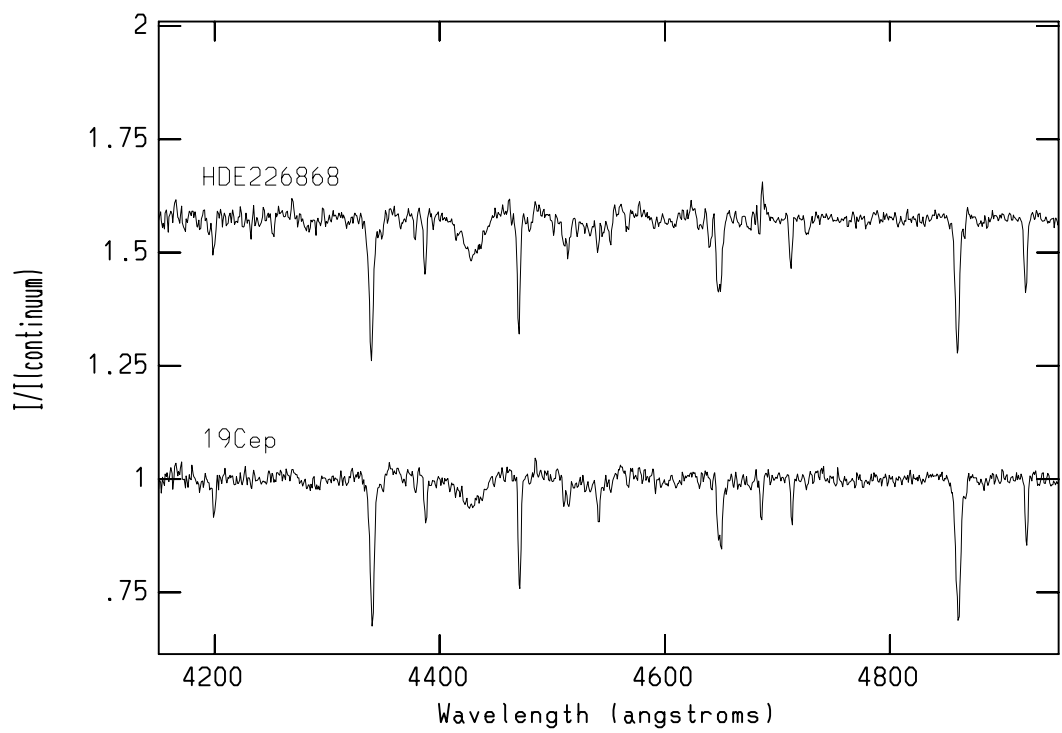
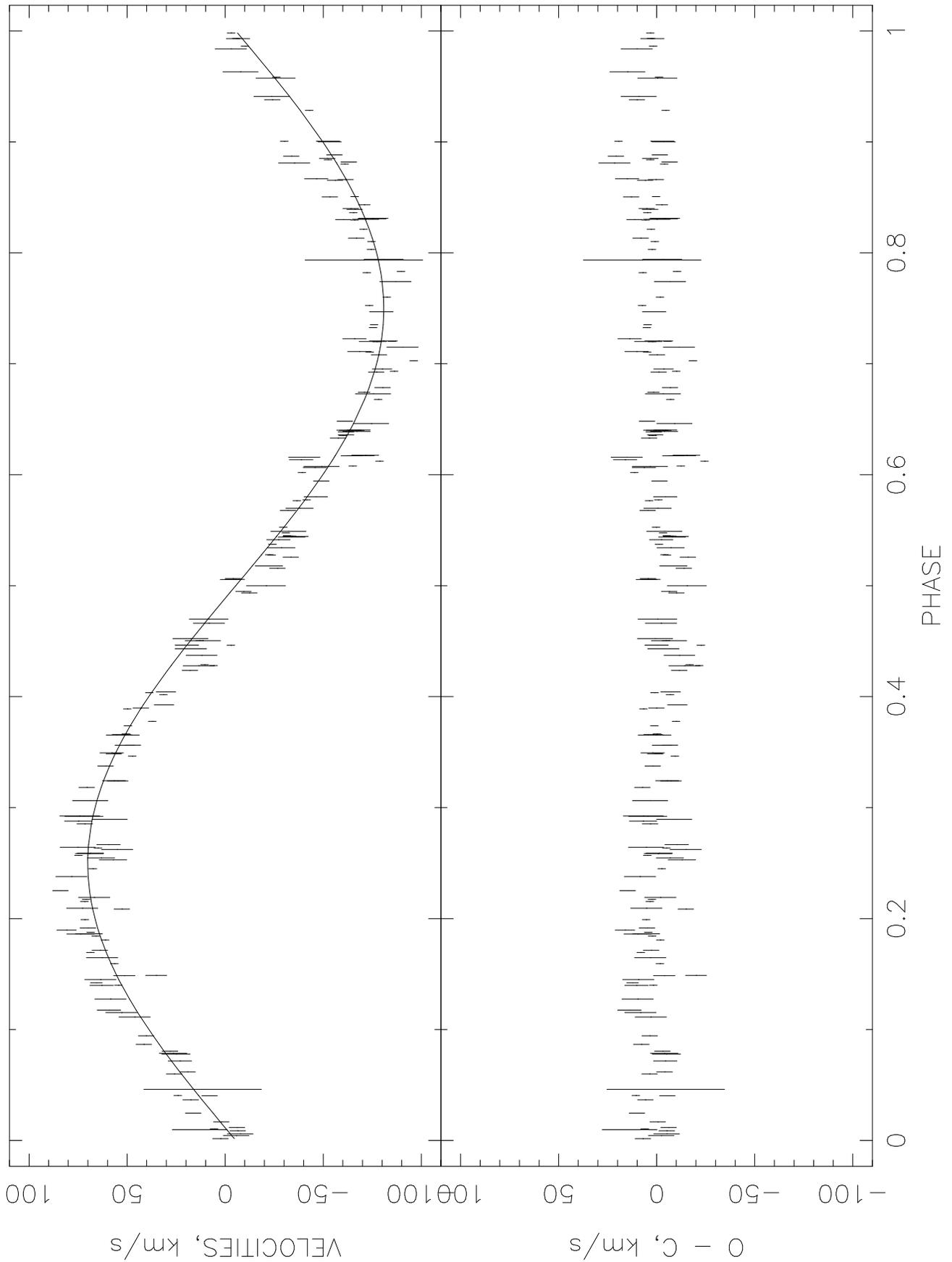
\end{document}